\documentclass[structabstract]{aa} %
\usepackage{graphicx} %
\usepackage{txfonts} %
\usepackage{natbib} %
\bibpunct{(}{)}{;}{a}{}{,} %
\begin{document}

\title{Dust-corrected surface photometry of \mbox{M\,31} \\from the Spitzer
  far-infrared observations}

\titlerunning{Dust-corrected photometry of \mbox{M\,31}}

\author{E.~Tempel\inst{1}\inst{2} \and A.~Tamm\inst{1} \and
  P.~Tenjes\inst{1}\inst{2}}

\offprints{E.~Tempel, \email{elmo@aai.ee atamm@ut.ee peeter.tenjes@ut.ee}}

\institute{Tartu Observatory, 61602 T\~oravere, Estonia \and Institute of
  Physics, Tartu University, T\"ahe 4, 51010 Tartu, Estonia}

\date{Received 26 March 2009 / Accepted 17 November 2009}

\abstract
{}
{We create a model for recovering the intrinsic, absorption-corrected surface
  brightness distribution of a galaxy and apply the model to the nearby galaxy
  \mbox{M\,31}.}
{We constructed a galactic model as a superposition of axially symmetric stellar
  components and a dust disc to analyse the intrinsic absorption effects. Dust
  column density is assumed to be proportional to the far-infrared flux of the
  galaxy. Along each line of sight, the observed far-infrared spectral energy
  distribution was approximated with modified black body functions corresponding
  to dust components with different temperatures, thereby allowing us to
  determine the temperatures and relative column densities of the dust
  components. We applied the model to the nearby galaxy \mbox{M\,31} using the
  Spitzer Space Telescope far-infrared observations for mapping dust
  distribution and temperature. A warm and a cold dust component were
  distinguished.}
{ The temperature of the warm dust in \mbox{M\,31} varies between 56 and 60\,K and
  is highest in the spiral arms, while the temperature of the cold component is mostly
  15--19\,K and rises up to about 25\,K at the centre of the galaxy. The
  intensity-weighted mean temperature of the dust decreases from $T \sim 32$\,K
  in the centre to $T \sim 20$\,K at $R\sim 7$\,\mbox{kpc} and outwards. The
  scalelength of the dust disc is $(a_0)_{\mathrm{dust}} \approx 1.8~
  (a_0)_{\mathrm{stars}}$. We also calculated the intrinsic $U$, $B$, $V$, $R$,
  $I$, and $L$ surface brightness distributions and the spatial luminosity
  distribution. The intrinsic dust extinction in the $V$-colour rises from
  $0.25^m$ at the centre to $0.4^m-0.5^m$ at $R\simeq 6-13$\,\mbox{kpc} and
  decreases smoothly thereafter. The calculated total extinction-corrected
  luminosity of \mbox{M\,31} is $L_B = (3.64 \pm 0.15) \cdot 10^{10}
  {L_{\sun}}$, corresponding to an absolute luminosity $M_B=-20.89\pm
  0.04$\,\mbox{mag}. Of the total $B$-luminosity, 20\,\% (0.24\,\mbox{mag}) is
  obscured from us by the dust inside \mbox{M\,31}. The intrinsic shape of the
  bulge is slightly prolate in our best-fit model.}
{}

\maketitle

\section{Introduction}

The importance of correcting astronomical observations for the impact of
intervening dust has been known for a long time. Several direct and indirect
methods have been developed for estimating light absorption and reddening by
interstellar matter. However, because of the complexity of this process, the
influence of dust is often ignored in studies of individual galaxies, as well as
extensive statistical samples. Unfortunately, ignoring dust effects seriously
undermines the reliability of the results, since it may lead to substantially
wrong estimates of the shapes, luminosities, colours, and masses of the stellar
components of galaxies \citep{Driver:07,padilla08, graham08}.

Two significantly different approaches can be used to estimate intrinsic dust
effects in external galaxies. Firstly, light extinction and reddening can be
traced directly by comparing the properties of affected and unaffected fields
within or behind a given galaxy. For example, in the presence of a partly
occulted background galaxy, the optical thickness and extinction curve of dust
inside the foreground galaxy can be measured directly
\citep{holwerda07}. Unfortunately, suitable pairs of galaxies are very
rare. Alternatively, extinction can be estimated by counting background galaxies
seen through the foreground galaxy and by comparing their abundance to
unobscured fields \citep{gonzalez98, gonzalez03, holwerda05}. The statistical
nature and difficulties in detecting galaxies behind brighter regions (e.g. the
bulge) of the foreground galaxy tend to reduce the applicability of this method.

The second approach exploits the law of conservation of energy -- optical and
ultra-violet radiation absorbed by dust has to be reradiated at the infrared
(\mbox{IR}) and longer wavelengths. Data collected by the Infrared Astronomical
Satellite (\mbox{IRAS}), the Infrared Space Observatory (\mbox{ISO}), the
Spitzer Space Telescope, and the Submillimetre Common-User Bolometer Array
(\mbox{SCUBA}) have significantly improved our understanding of the spatial
distribution, temperature, and physical properties of interstellar dust grains,
enabling the construction of radiation transfer models even down to the level of
individual dust grains. In addition to revealing new facts about cosmic dust
itself, these models enable restoration of the intrinsic photometric properties
of galaxies for which direct methods of estimating light extinction are not
applicable.

Since the thermal radiation from dust strongly depends on dust temperature,
far-\mbox{IR} emission maps taken at several wavelengths are required for
calculating dust density and for estimating the extinction of starlight. The
first proof that dust temperature may significantly vary within a galaxy was
noted from the \mbox{IRAS} data by \citet{Helou:86} and \citet{Lonsdale:87}. It
was suggested that, in addition to the warmer dust associated with star-forming
regions, a cooler and more diffuse dust component exists. Higher angular
resolution far-\mbox{IR} maps of several galaxies by the \mbox{ISO} satellite
indicated that the spatial distributions of the cold dust and the warm dust are
different \citep{haas98,Tuffs:03,Engelbracht:04,Hinz:04}. It was also discovered
that the cold dust disc is usually more extended than the stellar disc
\citep{Popescu:03}.

By approximating the \mbox{IRAS} and \mbox{SCUBA} measurements of dust spectral
energy distributions in 18 galaxies with two-component dust models,
\citet{Vlahakis:05} found that cold dust has temperatures 17--24\,K and warm
dust 28--59\,K. By using the Spitzer, \mbox{SCUBA} and \mbox{IRAS} observations
for a sample of 10 galaxies \citet{Willmer:09} derived the temperature range for
the cold dust 18--24\,K and for the warm dust 52--58\,K. The ratio of the cold
dust mass to the warm dust mass in these galaxies is typically 100--2000. These
values were derived by averaging across each of the galaxies; dust temperature
may be considerably different in specific regions.

To determine the total mass of dust in galaxies, \citet{Li:01},
\citet{Weingartner:01}, and \citet{Draine:07a} proposed a model considering
three kinds of dust grains and the radiation field, in which dust temperature
variations from galaxy to galaxy are regulated by the effective intensity of the
radiation field. This model has been successfully applied to 17 galaxies by
\citet{Draine:07} and more recently to 57 galaxies by
\citet{munoz-mateos}. \citet{Montalto:09} have used this model for studying
\mbox{M\,31}.

More sophisticated analyses, yielding dust temperature as a function of position
in a galaxy and thus enabling a detailed comparison to far-\mbox{IR} maps are
carried out within radiative transfer models of stellar radiation in a dusty
environment \citep[e.g.][]{Sauty:98,Silva:98,Bianchi:00,Popescu:00,Tuffs:04}.
Spatial resolution can reach 20\,\mbox{pc} in these models
\citep{Bianchi:08}. However, this is still not enough to resolve individual dust
clumps. The problem can be solved by separating the clumpy warm dust component
from the diffuse component, as done by \citet{Silva:98} and \citet{Popescu:00}.

In this paper, we create a three-dimensional galaxy model with axi-symmetric
stellar populations. To estimate extinction and to restore the intrinsic
luminosity and colour distributions of a galaxy, we add a dust disc to the
model. We apply the model to the nearby spiral galaxy \mbox{\object{M\,31}},
using the Spitzer far-\mbox{IR} maps (supplemented by the \mbox{IRAS}
observations) to determine the dust distribution. The distribution of the dust
temperature is estimated by approximating the far-\mbox{IR} spectral energy
distribution with modified black body functions assuming the presence of a
colder and a warmer dust component.

The following general parameters of M\,31 have been applied in our calculations:
the inclination angle has been taken 77.5\degr
\citep{walterbos88,devaucouleurs91}; the major axis position angle is 38.1\degr
\citep{walterbos87,ferguson02} and the distance to \mbox{M\,31} is
785\,\mbox{kpc} \citep{mcconnachie05}, corresponding to the scale
\hbox{1$\arcmin$ = 228\,\mbox{pc}}.

\section{Construction of the model}

The physical properties of dust and the extinction of the stellar light of a
galaxy are determined by an interplay between the spatial radiation field and
the dust grains at each location within the galaxy. Calculations of the
intrinsic extinction of a galaxy should thus be based on the spatial luminosity
distribution of the galaxy and the spatial density distribution of the dust
grains.

\subsection{Density distribution model for the stellar
  components}\label{sec:density_dist}

In this subsection, we develop a sufficiently flexible model for describing the
spatial distribution of the luminosity of a galaxy. A two-dimensional projection
of the model can be compared to the observed surface brightness distribution and
model parameters can be adjusted.

The model galaxy is given as a superposition of its individual stellar
components. Each visible component is approximated by an ellipsoid of rotational
symmetry with a constant axial ratio $q$; its spatial density distribution
follows the law
\begin{equation} l (a)=l (0)\exp \left[ -\left( {a \over ka_0}\right)^{1/N}
  \right] , \label{eq1}
\end{equation}
where $l (0)=hL/(4\pi q a_0^3)$ is the central density and $L$ is the component
luminosity; $a= \sqrt{r^2+z^2/q^2}$, where $r$ and $z$ are two cylindrical
coordinates; $a_0$ is the harmonic mean radius that characterises the real
extent of the component, independently of the structure parameter $N$. The
coefficients $h$ and $k$ are normalising parameters, dependent on $N$. The
definition of the normalising parameters $h$ and $k$ and their calculation is
described in appendix~B of \citet{tenjes94}. The luminosity density distribution
(\ref{eq1}) proposed by \citet{Einasto:65} is similar to the \citet{sersic68}
law for surface densities. Differences between the S\'{e}rsic law and the
two-dimensional projection of Eq.~\ref{eq1} can be seen in \citet{tamm05}.

Observations have demonstrated that the stellar disc can have a toroidal form in
some galaxies, i.e. it does not continue to the centre. As suggested by Einasto
\citep{einasto69,einasto80}, the spatial density of a disc with a central hole
can be expressed as a sum of two spheroidal mass distributions
\begin{equation} \label{eq:lumhole} l_{\mathrm{disc}}(a) = l_{+}(a) + l_{-}(a) ,
\end{equation}
both of which can be approximated with the exponential law (\ref{eq1}). Adopting
a model disc with zero density at $r=0$ and non-negative density at
$l_{\mathrm{disc}}(a) > 0$ gives the following relations between the parameters
of the components $l_{+}$ and $l_{-}$: $a_{0-}=\kappa a_{0+}$,
$L_{-}=-\kappa^2L_{+}$, $q_{-}=q_{+}/\kappa$, where $\kappa < 1$ is a parameter
that determines the relative size of the hole in the centre of the
disc. Structural parameters $N_{-}$ and $N_{+}$ should be identical.

Density distributions of all visible components are projected along the line of
sight and their sum yields the surface brightness distribution of the model
\begin{equation} L(A)= 2\sum_{j}\frac{q_j}{Q_j} \int\limits_A^\infty
  \frac{l_j(a)a\,\mathrm{d}a}{(a^2-A^2)^{1/2}} , \label{eq2}
\end{equation}
where $A$ is the major semi-axis of the equidensity ellipse of the projected
light distribution and $Q_j$ are their apparent axial ratios
$Q^2=\cos^2i+q^2\sin^2i$. The inclination angle between the plane of the galaxy
and the plane of the sky is denoted by $i$. The summation index $j$ designates
each visible component.

To approximate the luminosity distribution of a given galaxy, an arbitrary
number of components can be used. Obviously, a larger number of components
enables a better approximation of observations; however, to keep the model
physically justified, additional information (e.g. observations of metallicity
and kinematics distribution) should be used for distinguishing different
components and for determining their number and extent. For the majority of
nearby regular galaxies, three to five stellar components with density
distribution (\ref{eq1}) should be enough for describing the luminosity
distribution.

The model parameters of each component are determined by a least-squares
approximation of the observed surface brightness distributions with the model
profiles via Eq.~\ref{eq2}. Uncertainties of the model can be estimated by using
the partial second derivatives of the sum of the least-square differences
\citep{bevington03}.

\subsection{The dust disc component}\label{sec:dustdisc}

\begin{figure}
  \resizebox{\hsize}{!}{\includegraphics{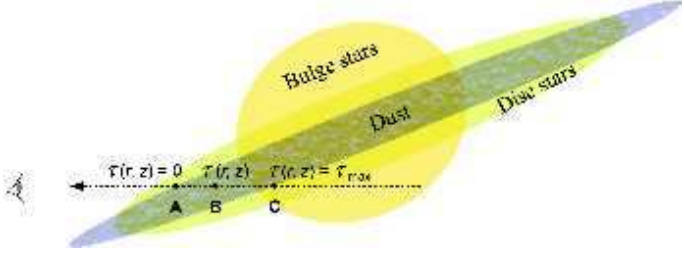}} \caption{Geometry of the
    dust disc model. The line-of-sight optical depth for three positions A, B,
    and C is indicated.  The figure is illustrative
    only.} \label{fig:extinction}
\end{figure}
Let us now implant a dust disc component into our galaxy model, allowing us to
calculate absorption along each line of sight through the galaxy.

Equation~\ref{eq2} will give the luminosity density along a line of sight, if no
dust extinction is included. The intrinsic luminosity density for each component
can be calculated from the equation
\begin{equation} L(X,Y)=\int\limits^{\infty}_{X}
  \frac{\sum\limits_{j=1}^2\left[l(r,z_j)\,\mathrm{e}^{-\tau(r,z_j)}\right]}
  {\sin{i}\,\sqrt{r^2-X^2}}r\,\mathrm{d}r,\label{eq:dust_lumin}
\end{equation}
\begin{equation} z_{1,2}= \frac{Y}{\sin{i}} \pm \frac{\sqrt{r^2-X^2}}{\tan{i}},
\end{equation}
where $l(r,z)$ denotes the spatial luminosity density (\ref{eq1}) of the galaxy
component, $L(X,Y)$ is the corresponding surface brightness distribution, $X$
and $Y$ are coordinates in the plane of the sky, and $\tau(r,z)$ is the optical
depth along the line of sight (see Fig.~\ref{fig:extinction}).

Optical depth $\tau(r,z)$ is zero between the observer and the dust disc and
$\tau(r,z)=\tau_{\mathrm{max}}$ behind the dust disc. $\tau_{\mathrm{max}}$ is
the total optical depth along a given line of sight and thus a function of
$(X,Y)$. Inside the dust disc, $\tau(r,z)$ is between $0$ and
$\tau_{\mathrm{max}}$.

The optical depth between the observer and each point inside the dust disc can
be written as
\begin{equation} \label{eq:dust_gen} \tau(r,z)=\tau_{\mathrm{max}} (X,Y)
  \frac{\int\limits^{\mathrm{B}}_{\mathrm{A}}n_{\mathrm{dust}}(s)\,
    \mathrm{d}s}{\int\limits^{\mathrm{C}}_{\mathrm{A}}n_{\mathrm{dust}}(s)\,\mathrm{d}s},
\end{equation}
where integration is done along the line of sight $s$ and $n_{\mathrm{dust}}(s)$
is the spatial number density of the dust disc.

To gain a simple but nevertheless flexible model for the dust disc, we adopt the
following form for $n_{\mathrm{dust}}$:
\begin{equation}
  \label{eq:ntau}
  n_{\mathrm{dust}}(r,z)=f_{\mathrm{dust}}(r)\cdot n (a),
\end{equation}
where $n (a)$ is the exponential law (\ref{eq1}) and $f_{\mathrm{dust}}(r)$
describes deviations from the exponential law (\ref{eq1}). The spatial number
density of dust decreases exponentially in the $z$-direction according to
Eq.~\ref{eq1}. Multiplication by $f_{\mathrm{dust}}(r)$ allows density behaviour
to be more flexible along the $r$-direction.

Assuming that extinction is proportional to the dust column density and the
latter is proportional to the far-\mbox{IR} flux, the parameters of $n(a)$, the
shape of $f_{\mathrm{dust}}(r)$ and subsequently the map of
$\tau_{\mathrm{max}}(X,Y)$ can be derived. The relation between reddening and
the far-\mbox{IR} flux can be expressed as
\begin{equation}
  \label{eq:reddeningMW}
  E(B-V) = pD,
\end{equation}
where $D$ is a measure of the dust column density and $p$ is a calibration
constant \citep{schlegel98}. Expressing $D$ as the 100\,\mbox{$\mu$m} flux of
dust with a reference temperature 18.2\,K, \citet{schlegel98} have estimated
$p=0.016\pm0.004$ by studying the colours of bright elliptical galaxies as seen
through the Galactic dust.

To make Eq.~\ref{eq:reddeningMW} applicable for other galaxies, we need to take
in account the difference between the 100\,\mbox{$\mu$m} flux of dust at the
temperature 18.2\,K and the actual flux of dust with a temperature $T$ at a
wavelength $\lambda$. According to the modified Planck function
\begin{equation}
  \label{eq:dustcoldens}
  D(X,Y)=F_{\lambda}(X,Y) \frac{\lambda^\beta B_\lambda(100\,\mu\mathrm{m},18.2\,\mathrm{K})}
  {(100\,\mu\mathrm{m})^\beta B_\lambda(\lambda,T(X,Y))},
\end{equation}
where $F_{\lambda}(X,Y)$ is the measured far-\mbox{IR} flux map at a wavelength
$\lambda$ and $T(X,Y)$ is the temperature map, $B_\lambda$ is the black body
function and $\beta$ is the dust emissivity index. According to
\citet{Vlahakis:05} we take $\beta=2.0$.

Using Eq.~\ref{eq:reddeningMW} and the relation $A_V=R\cdot E(B-V)$ and
converting the derived extinction into optical depth, we reach a map of the
total optical depth $\tau_{\mathrm{max}}(X,Y)$
\begin{equation}
  \label{eq:taumax}
  \tau_{\mathrm{max},\lambda}(X,Y)=\frac{\ln(10)}{2.5}\frac{A_\lambda} {A_V}RpD(X,Y).
\end{equation}
\citet{Xilouris:99} have found that in several galaxies, the extinction law
agrees well with that of the Milky~Way. This is also the case for \mbox{M\,31}
as indicated by the colours of its globular clusters \citep{barmby00}. We can
thus use $p=0.016$ and $R=3.1$; the extinction ratio $A_\lambda/A_V$ can be
determined from the Galactic extinction law.

Equation~\ref{eq:taumax} gives the total extinction map in the plane of the
sky. To calculate the actual intrinsic extinction, we also need the spatial
density distribution of the dust disc $n_{\mathrm{dust}}(r,z)$ for
Eq.~\ref{eq:dust_gen}. We can compare the projection of the model space density
distribution (calculated according to Eq.~\ref{eq2}) to the observed dust column
density distribution (i.e. the far-\mbox{IR} maps) to determine the function
$f_\mathrm{dust}(r)$ and the parameters for $n(a)$. The observed column density
of the dust disc is thereby converted into its spatial number density. More
precisely, we can only determine the shape of the number density distribution of
the dust disc; its absolute calibration $n(0)$ remains unknown. In the present
case, the calibration constant is cancelled out in Eq.~\ref{eq:dust_gen} and is
not required for our model.

In the final step, the derived optical depth map
$\tau_{\mathrm{max},\lambda}(X,Y)$ and the function $n_{\mathrm{dust}}(r,z)$ are
inserted into Eq.~\ref{eq:dust_gen} and the extinction-corrected surface
brightness along each line of sight can be calculated from
Eq.~\ref{eq:dust_lumin}.

The relation between $\tau_{\mathrm{max},\lambda}(X,Y)$ and reddening in
Eq.~\ref{eq:taumax} is calibrated according to the reddening of extragalactic
sources by the dust of the Milky Way, thus including both extinction and
scattering. On the other hand, the luminosities calculated from
Eq.~\ref{eq:dust_lumin} include dust attenuation, but ignore the scattering of
photons, leading to an overestimation of stellar emission in regions with high
dust density. However, we consider the additional uncertainties caused by
scattering to be lower than other potential error sources in $\tau$
determination.

\section{Applying the model to \mbox{M\,31}}
\subsection{Optical data}

A large collection of surface brightness distribution measurements is available
for modelling the stellar luminosity distribution in \mbox{M\,31}; see
\citet{tenjes94} for references to earlier data. For the present work we have
used the measurements of $U$, $B$, $V$, $R$, and $I$ distributions excluding
some of the oldest, mainly photometrical data; however, for a more complete
coverage of $B$ surface brightness distribution, we have also included the
photoelectrical measurements by \citet{de-Vaucouleurs:58}. As an example of
nearly absorption-free observations, we have added the 3.6\,\mbox{$\mu$m}
measurements by the Spitzer Space Telescope \citep{Barmby:06} to our dataset,
which roughly correspond to the Johnson-Cousins $L$ filter. Since we did not
intend to model the nuclear part of \mbox{M\,31}, data at $r < 0.3$\,\mbox{kpc}
were neglected. Table~\ref{photom} presents references to the observations and
the corresponding filter systems used in this paper. Some of the data are
one-dimensional measurements along either or both of the axes of the galaxy,
others are elliptically averaged contours of the galaxy. Measurements along the
axes are useful for an accurate determination of the three-dimensional shape of
the stellar populations, while ellipses provide the average surface brightness
at each distance from the centre. Of the minor axis data, only those from the
side closer to us (the southeast direction) have been used and are presented on
figures throughout this paper. Different $R$ and $I$ colour system profiles were
transferred into the Johnson-Cousins system, using the calibrations by
\citet{frei:94}. All measurements were corrected for extinction by dust of the
Milky~Way according to \citet{schlegel98}.
\begin{table} \caption{Used photometrical data.} \label{photom}
  \centering \begin{tabular}{lll} \hline\hline Reference & Filter & Remarks \\ &
    system & \\ \hline \citet{de-Vaucouleurs:58} & $B$ & ellipses \\
    \citet{Hoessel:80} & $U\/B\/V$ & major and \\ & & minor axis\\
    \citet{Hodge:82} & $B$ & ellipses \\ \citet{Kent:87} & $R$ & ellipses \\
    \citet{walterbos88} & $U\/B\/V\/R$ & ellipses \\ \citet{Pritchet:94} & $V$ &
    star counts\\ & & (minor axis) \\ \citet{Guhathakurta:05} & $V$ & star
    counts\\ & & (minor axis) \\ \citet{irwin05} & $V\/I$ & ellipses and \\ & &
    star counts\\ & & (minor axis) \\ \citet{Barmby:06} & $L$ & ellipses \\
    \noalign{\smallskip} \hline \end{tabular} \begin{list}{}{} \item[] Note:
    `ellipses' denote elliptically averaged surface brightness
    profiles. \end{list}
\end{table}

\subsection{Infrared data}

Our analysis of the far-\mbox{IR} emission from \mbox{M\,31} is based on the
highest spatial resolution mappings currently available, the Spitzer Space
Telescope \mbox{MIPS} camera imaging at 24, 70, and 160\,\mbox{$\mu$m}
\citep{gordon06}, supplemented by the \mbox{IRAS} satellite detections at
100\,\mbox{$\mu$m}. The only measurements available at longer wavelengths are
those by the \mbox{DIRBE} instrument aboard the \mbox{COBE} satellite at up to
240\,\mbox{$\mu$m}. Because of their very low angular resolution, the
\mbox{DIRBE} observations were not suitable for this work.

The Spitzer images were downloaded using the Leopard toolkit developed at the
Spitzer Science Center. An inspection of the pipeline-created (post-\mbox{BCD})
mosaic images revealed uneven background levels and the presence of a number of
instrumental defects, making them unusable for our purposes. We started from the
basic-calibrated data (\mbox{BCD}) and conducted the necessary corrections,
stacking and mosaicing using the standard procedures within the \mbox{IRAF} and
\mbox{MOPEX} software. The first three frames of each 24\,\mbox{$\mu$m}
observing run, suffering from a shorter exposure time, were discarded. Depending
on the fields, either a constant or a low order polynomial background was
subtracted from the images. The \mbox{IRAS} imaging was retrieved from the
\mbox{NASA/IPAC} Infrared Science Archive. The resolution of the \mbox{IRAS}
data are only 90\,arcsec/pixel; they were not used during the final
approximation of the spectral energy distribution. The final images were sampled
to 8\,arcsec/pixel, which is the sampling of the 160\,\mbox{$\mu$m}
observations; this also became the spatial resolution of our dust model.
\begin{figure}
  \resizebox{\hsize}{!}{\includegraphics{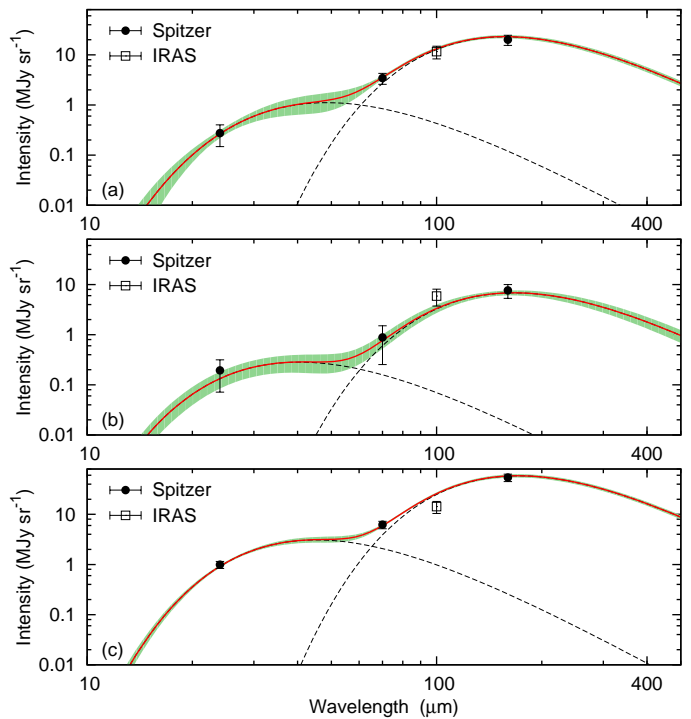}} \caption{Examples
    of approximations of the observed Spitzer (filled circles) and \mbox{IRAS}
    (empty squares) data with the modified black body functions at three random
    locations: (a)~in the bulge region; (b)~in an interarm region; (c)~in a
    spiral arm. In each panel, the black dashed lines represent the warm and the
    cold component and the red solid line is the sum of the two modified black
    body functions together with its uncertainty
    region.} \label{fig:dustmodel_pts}
\end{figure}

Uncertainties were estimated considering three principal error sources: the
characteristics of the telescope detectors, the precision of the absolute
calibration and the uncertainty of the background level. An estimate for the
response uncertainty of each detector pixel is indicated by the corresponding
standard deviation maps; these were stacked and resampled by the \mbox{MOPEX}
software during the mosaicing process. The conservative estimates of the
uncertainties of the instrumental flux calibration of Spitzer were taken from
\citet{engelbracht07}, \citet{gordon07}, and \citet{stansberry07}: 4\,\%, 7\,\%,
and 12\,\% for 24, 70, and 160\,\mbox{$\mu$m}, respectively. \citet{beichman88}
give 2\,\% as the maximal variation in the \mbox{IRAS} absolute
calibration. Possible background errors were estimated from deviations between
the background flux density levels of neighbouring data stripes. Combining these
three error sources, uncertainty maps were produced for each image, to be later
used for calculating possible errors in our extinction analysis. In addition,
the uncertainty of disentangling the contributions of \mbox{M\,31} and the
Milky~Way from the far-\mbox{IR} maps has to be kept in mind. According to the
\mbox{IRAS} maps, the sky brightness at 100\,\mbox{$\mu$m} around \mbox{M\,31}
varies between 2 and 5\,\mbox{$\mathrm{MJy\,sr^{-1}}$} \citep{schlegel98}. This
also gives an idea about a possible error resulting from wrong addressing of
dust in the direction of \mbox{M\,31}. We have not included this foreground
uncertainty in our error estimates.
\begin{figure}
  \resizebox{\hsize}{!}{\includegraphics{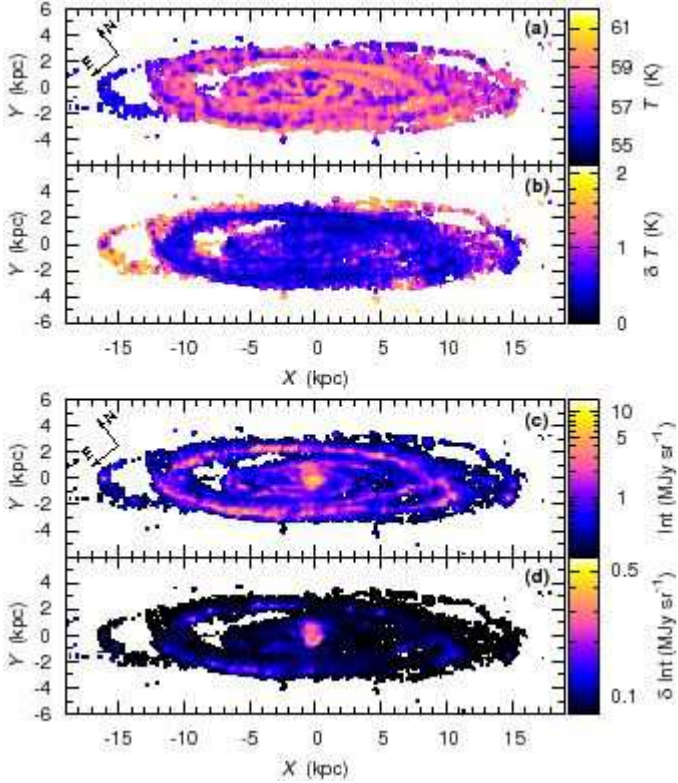}} \caption{Temperature
    and intensity maps of the warm dust component. In panels (a) and (b) the
    warm dust temperature and its $1\sigma$-errors are given. In panels (c) and
    (d) the intensity of the warm component at 100\,\mbox{$\mu$m} and its
    $1\sigma$-errors are given.} \label{fig:tolm1}
\end{figure}
\begin{figure}
  \resizebox{\hsize}{!}{\includegraphics{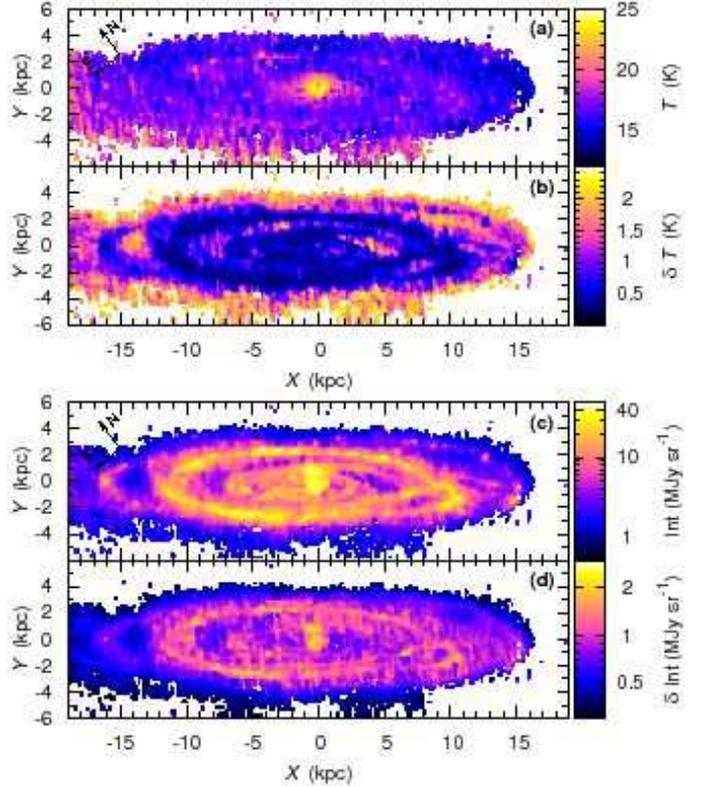}} \caption{Temperature
    and intensity maps of the cold dust component. In panels (a) and (b) the
    cold dust temperature and its $1\sigma$-errors are given. In panels (c) and
    (d) the intensity of the cold component at 100\,\mbox{$\mu$m} and its
    $1\sigma$-errors are given.} \label{fig:tolm2}
\end{figure}

To probe the repeatability of our image processing and photometry, let us
compare our estimate for the integral \mbox{IR} flux to some earlier
measurements. The total sky-subtracted flux within the outermost ellipse fitted
to the galaxy with the \mbox{IRAF/STSDAS} task \emph{ellipse\/} is
119$\pm$7\,\mbox{Jy}, 1200$\pm$110\,\mbox{Jy}, and 7800$\pm$1000\,\mbox{Jy} for
the 24, 70, and 160\,\mbox{$\mu$m} images, respectively. These correspond rather
well to the values 107$\pm$10\,\mbox{Jy}, 940$\pm$188\,\mbox{Jy}, and
7900$\pm$1580\,\mbox{Jy} measured by \citet{gordon06} inside a rectangular
aperture (2.75\degr $\times$ 0.75\degr). The variations probably come from
different methods used for background subtraction and from the aperture
selection.

\subsection{Mapping dust temperature and optical depth}

Following Eqs.~\ref{eq:dustcoldens} and~\ref{eq:taumax}, we need a far-\mbox{IR}
emission map and a map of the dust temperature. Since we are relying on the
relation between reddening and the far-\mbox{IR} flux at 100\,\mbox{$\mu$m}
determined by \citet{schlegel98} for the Milky~Way, the far-\mbox{IR} map has to
correspond to the 100\,\mbox{$\mu$m} flux. Although the \mbox{IRAS} satellite
has observed at this wavelength, we derived the necessary map by interpolating
the Spitzer observations because of their higher spatial resolution.

\citet{gordon06} have shown that a two-component dust model gives a good fit to
the far-\mbox{IR} spectral energy distribution of \mbox{M\,31}. The warmer (T
$\approx$ 59\,K) dust component is about $10^4$ times less massive than the
colder (T $\approx$ 17\,K) component, but it might still cause significant light
extinction in specific regions, since it is considerably more clumpy and more
concentrated in the galactic centre and in the spiral arms. Therefore, we have
fitted the far-\mbox{IR} emission within each line of sight with two modified
Planck functions, representing these two dominant temperature components.

To determine the intensity and temperature of both of the dust components, the
modified Planck functions were fitted to the 4-bin spectral energy distribution
observations (24, 70, 100, and 160\,\mbox{$\mu$m}) in three stages for each line
of sight (i.e. within each model pixel).  The main contribution to the warm
component comes from the star-forming regions. We assumed that the temperature
of dust in these regions varies only slightly and fixed the temperature of the
warm component at $T=59$\,K \citep{gordon06} during the first fitting stage; the
other parameters were derived using the Spitzer data. During the second stage,
the IRAS data were added and the temperature of the warm dust was adjusted at
lower spatial resolution. Finally, \mbox{IRAS} data were omitted again and the
intensities of the dust components and the temperature of the cold dust were
refined at higher spatial resolution. The uncertainties of the derived
intensities and temperatures were estimated by using the Monte Carlo method,
allowing the data points to vary randomly within their individual
uncertainties. For each model pixel, 100 different sets of data points were
created and approximated with the two-component model. Examples of the black
body fits to the data points for three randomly selected lines of sight
(i.e. for three model pixels) are given in Fig.~\ref{fig:dustmodel_pts}. The
final temperature maps and 100\,\mbox{$\mu$m} intensity maps of the two dust
components are presented in Figs.~\ref{fig:tolm1} and~\ref{fig:tolm2} together
with the corresponding 1$\sigma$-error maps.
\begin{figure}
  \resizebox{\hsize}{!}{\includegraphics{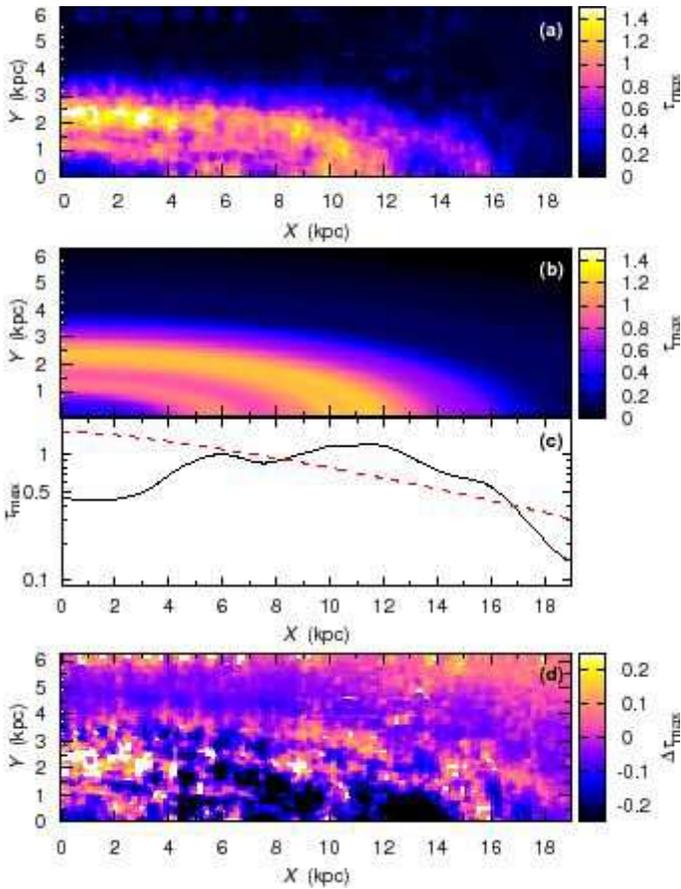}} \caption{Four-folded
    optical depth map of \mbox{M\,31}, derived from the cold dust component
    using Eq.~\ref{eq:taumax}. Panel~(a)~-- the observed optical depth;
    panel~(b)~-- the modelled optical depth along the line of sight;
    panel~(c)~-- a cut from panel~(b) for $Y=0$ (black solid line); the red
    dashed line corresponds to the dust disc model with a simple exponential
    density distribution law (\ref{eq1}); panel~(d)~-- the residual map of the
    observed and the modelled optical depth.} \label{fig:tolm3}
\end{figure}

Next, the total optical depth $\tau_{\mathrm{max}}$ (applicable to light sources
behind the dust disc) was calculated for each model pixel according to
Eqs.~\ref{eq:dustcoldens} and~\ref{eq:taumax}.  Only the cold dust component was
used for calculating the optical depths, thus the calculated opacities refer to
the diffuse dust component. Since the mass of the warm dust component is very
low (see Sect.~\ref{sec:results}), its contribution to the optical depth would
be negligible. The derived $\tau_{\mathrm{max},V}(X,Y)$ map was four-folded (see
Fig.~\ref{fig:tolm3}a) and approximated with an axially symmetric distribution
(\ref{eq:ntau}) as described in Sect.~\ref{sec:dustdisc} (see
Fig.~\ref{fig:tolm3}b). The residual map, indicating deviations between the
axially symmetric model and the four-folded map is shown in
Fig.~\ref{fig:tolm3}d.

Approximating the far-\mbox{IR} maps with the axially symmetric distribution
(\ref{eq:ntau}) does no provide an independent estimate for the vertical
thickness of the dust disc component. We have used a rather conventional
assumption that the mean ratio of the thickness of the stellar disc to the
thickness of the dust disc is $1.8 \pm 0.6$, derived by modelling a sample of
nearby edge-on galaxies by \citet{Xilouris:99}. However, in some galaxies the
dust disc appears thinner \citep{Bianchi:07}, thus we have made comparative
calculations also for the cases of a thinner and a thicker dust disc (see
Sect.~\ref{sec:discussion} for models with different dust disc thicknesses). The
scalelength of the dust disc was determined during the approximation process of
the far-\mbox{IR} emission maps. The structural parameters for the exponential
approximation of the dust disc density distribution $n(a)$ are given in
Table~\ref{param}. The dust disc density profile (\ref{eq:ntau}) in the
$r$-direction is given in Fig.~\ref{fig:tolm3}c, also showing that without a
correction function $f_{\mathrm{dust}}(r)$, the simple exponential function
$n(a)$ provides a very poor approximation.
\begin{figure}
  \resizebox{\hsize}{!}{\includegraphics{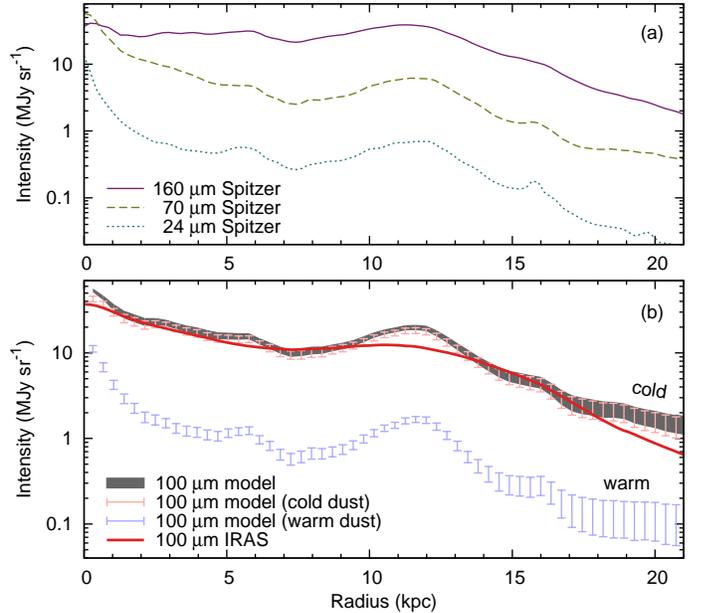}} \caption{Far-\mbox{IR}
    intensity distribution of \mbox{M\,31}, derived by fitting ellipses to the
    Spitzer, \mbox{IRAS}, and modelled data. Upper panel~-- the Spitzer 24, 70,
    and 160\,\mbox{$\mu$m} intensity distributions. Lower panel~-- the modelled
    intensity at 100\,\mbox{$\mu$m} for the cold and warm components (red and
    blue errorbars), and their weighted average (filled area); the red solid
    line represents the \mbox{IRAS} 100\,\mbox{$\mu$m} data.} \label{ir_vrd}
\end{figure}

\subsection{Model fitting}

\begin{table*}
  \caption{Calculated parameters of the photometrical model.}
  \label{param}
  \centering
  \begin{tabular}{lllllll}
    \hline\hline
    Population & $a_0$ & $q$ & $N$ & $\kappa$ & $k$ & $h$ \\
    & (kpc) & & & & & \\
    \hline
    Bulge &0.50$\pm0.01$&1.06$\pm0.03$&3.46$\pm0.11$ &&$7.2086\cdot10^{-4}$&896.897
    \\
    Disc & 6.7$\pm0.2$ &0.08$\pm0.01$&1.90$\pm0.13$ &&$6.4723\cdot10^{-2}$&26.7655
    \\
    Young disc$^{(a)}$& 9.5$\pm0.2$ &0.02$\pm0.003$&0.35$\pm0.02$&0.67$\pm0.02$&1.3334& 1.23802
    \\
    Halo & 6.5$\pm0.4$ &0.50$\pm0.02$&5.0$\pm0.2$ &&$4.1625\cdot10^{-6}$&31809.6
    \\
    Dust disc & 17.0 &0.014 &1.0 &&0.5000&4.00000
    \\
    \hline
    Population & $L_U$ & $L_B$
    & $L_V$ & $L_R$
    & $L_I$& $L_L$ \\
    & ($10^{10}\mathrm{L_{\sun}}$) &
    ($10^{10}\mathrm{L_{\sun}}$) & ($10^{10}\mathrm{L_{\sun}}$) &
    ($10^{10}\mathrm{L_{\sun}}$) & ($10^{10}\mathrm{L_{\sun}}$)&
    ($10^{10}\mathrm{L_{\sun}}$) \\
    \hline
    Bulge & 0.63$\pm0.07$& 0.82$\pm0.04$& (0.96$\pm0.05$) & (0.93$\pm0.04$) &
    0.83$\pm0.09$ & 4.08$\pm0.22$ \\
    Disc & 0.61$\pm0.07$& 0.90$\pm0.06$& 1.36$\pm0.07$ & 1.47$\pm0.13$ &
    (1.88$\pm0.27$) & 5.40$\pm0.52$ \\
    Young disc$^{(b)}$& 0.98$\pm0.11$&1.10$\pm0.08$ &0.85$\pm0.08$ &0.83$\pm0.13$ &
    (1.19$\pm0.21$) & 1.09$\pm0.58$ \\
    Halo & (0.79$\pm0.19$)& (0.82$\pm0.11$)& 0.23$\pm0.02$ & (1.70$\pm0.28$) &
    0.72$\pm0.06$ & (2.32$\pm0.95$) \\
    \hline
  \end{tabular}
  \begin{list}{}{}
  \item[] Notes: The luminosities are corrected for the intrinsic absorption and
    the absorption in the Milky~Way. The luminosities supported by insufficient data
    are given in brackets. $^{(a)}$~Structural parameters for the young disc are
    only given for the positive component; see Sect.~\ref{sec:density_dist} for
    determining the parameters of the negative component. $^{(b)}$~The
    luminosities of the young disc are its true luminosities ($L=L_{+}+L_{-}$).
  \end{list}
\end{table*}

Prior to starting the actual fitting process, a reasonable number of different
stellar components to be incorporated in the model was specified.

The two most dominant components of \mbox{M\,31} are its bulge and disc,
detectable even with an amateur telescope. In addition, recent studies suggest
the presence of a dynamically warm but also slowly rotating stellar population
of stars with moderate metallicity beginning at the outskirts of the bulge
region. Here we designate this population as the halo; note that the same region
has been referred to as the bulge \citep{kalirai06}, the extended disc
\citep{ibata05, ibata07}, the spheroid \citep{brown07}, and the outer halo
\citep{durrell01}, depending on the viewpoint and the focus of each
study. Furthermore, recent observational studies of distant fields in the
direction of the minor axis have provided evidence for the existence of stellar
populations far beyond the conventional bulge and halo regions \citep{irwin05,
  ibata07}. The outermost regions seem to be dominated by a faint, diffuse, very
extended metal poor halo, possibly embedding much of the whole Local
Group. Since it has almost negligible surface brightness and total luminosity,
we do not model this population as a separate component.

Also, we made no attempt to include the Giant Stream or any other of the
recently detected stellar streams in our model separately. Their contribution to
the surface brightness, either measured along the axes of the galaxy or averaged
along elliptical contours is insignificant.

Consequently, the fitting process was started with three stellar components: a
bulge, a disc, and a halo. However, we soon realised that an additional
component is necessary for a better description of the blue star-forming ring at
about 10\,\mbox{kpc} from the centre, also coinciding with a peak in the
distribution of gas clouds and other objects typical of regions of star
formation \citep{einasto80, tenjes91, tenjes94}. The importance of considering
clumpy young populations in models of the intrinsic dust extinction of galaxies
is also stressed in radiative transfer models \citep[e.g.][]{Bianchi:08}. We
inserted the young stellar population of \mbox{M\,31} in our model as a disc
with a central hole described by Eq.~\ref{eq:lumhole}; we refer to this
population as the `young disc' in the following.

The model parameters $q$, $a_0$, $L$, $N$, and $\kappa$ for each component were
determined by a subsequent least-squares approximation process. First, we made a
crude estimation of the population parameters. The purpose of this step is to
avoid obviously non-physical parameters -- relation (\ref{eq:dust_lumin}) is
non-linear and fitting a model to the observations is not a straightforward
procedure. Next, a mathematically correct solution was found for each component.
Details of the least-squares approximation and the general modelling procedure
have been described by \citet{einasto89}, \citet{tenjes94}, and
\citet{tenjes98}. Uncertainties of the model due to observational uncertainties
and the coupling of parameters were estimated using the partial second
derivatives of the sum of the least-square differences \citep{bevington03}. For
each parameter, the calculated errors only reflect the goodness of the given
fit; they do not represent the uniqueness of the model. Models with considerably
different parameters can be fitted to the observational data, yielding somewhat
larger deviations.

\section{Results}
\label{sec:results}

\begin{figure*}
  \centering \includegraphics[width=180mm]{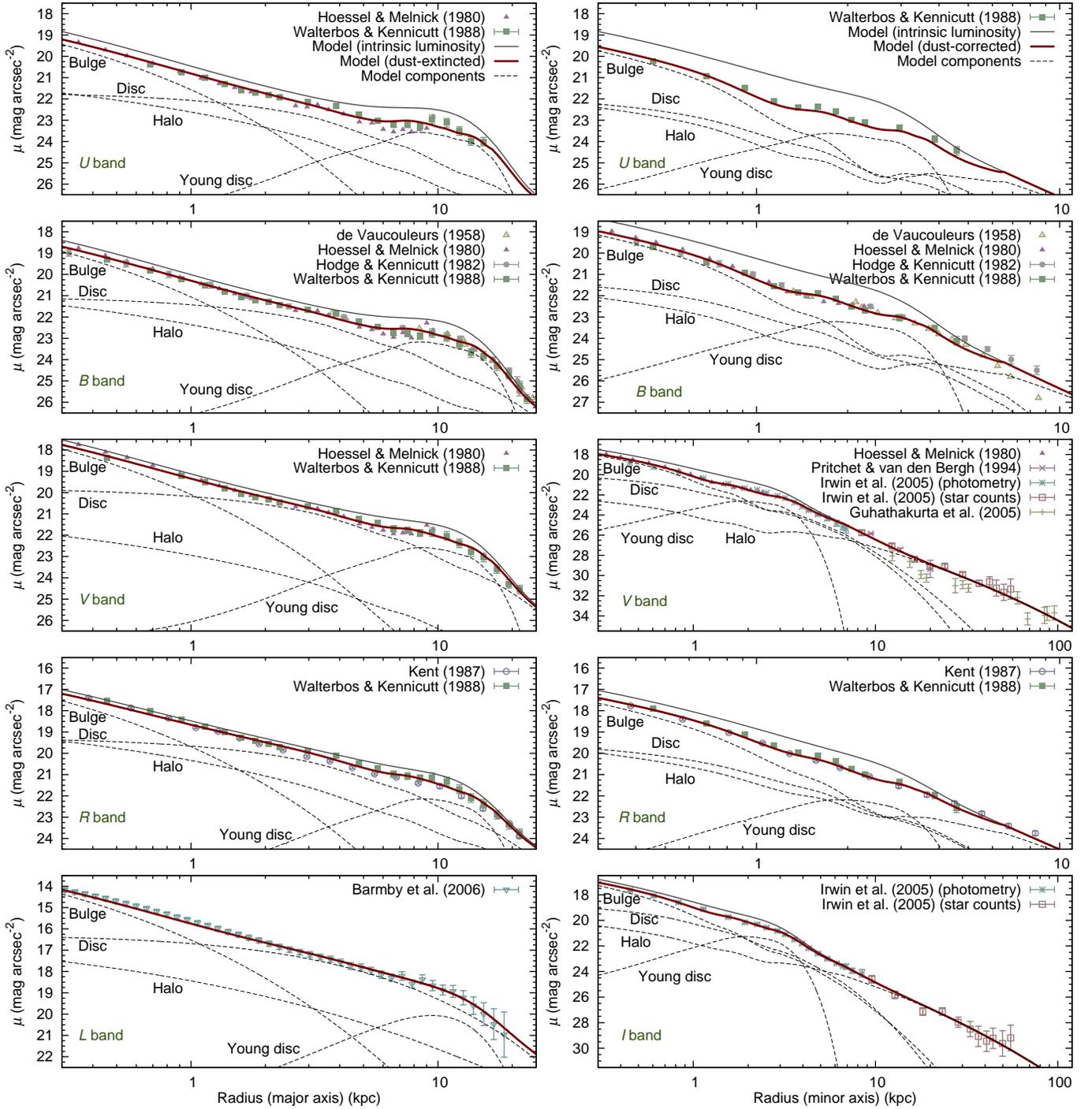} \caption{The observed
    (data points) and the modelled, dust-extincted (thick red solid lines)
    surface brightness distributions of \mbox{M\,31} together with the
    contributions by the individual stellar components (dashed lines). Thin
    solid lines represent the intrinsic, dust-free surface brightness
    distributions. Left panels: $U\/B\/V\/R\/L$ (from top to bottom) surface
    brightness along the major axis; right panels: $U\/B\/V\/R\/I$ (from top to
    bottom) surface brightness along the minor axis.} \label{heledus}
\end{figure*}

In the presented model, the dust disc of \mbox{M\,31} has been assumed to
consist of a cold component and a warm component. The temperature and intensity
distribution maps together with their $1\sigma$-errors for the warm component
are given in Fig.~\ref{fig:tolm1}. It is seen that the temperature of the warm
dust is 56--60\,K in most cases, with a typical uncertainty less than 1\,K. The
highest temperatures and intensities can be found in the spiral arms, revealing
a relation between the warm dust and the star-forming
regions. Figure~\ref{fig:tolm2} shows the temperature and intensity distribution
maps of the cold dust. Temperature of the cold component is usually 15--19\,K
with an uncertainty 0.2--1.5\,K. In contrast to the warm dust, the temperature
of the cold dust rises towards the centre of the galaxy, up to 25\,K. This can
be explained by a stronger general radiation field resulting from the higher
concentration of stars. The intensity distribution of the cold component is much
smoother than that of the warm component, matching well with its more diffuse
nature. The integral dust temperatures estimated by \citet{gordon06}, 17\,K for
the cold component and 59\,K for the warm component, agree well with the average
values derived in our study.

We have used two-dimensional far-\mbox{IR} maps of \mbox{M\,31} for our
extinction analysis. For illustration, Fig.~\ref{ir_vrd}a shows the Spitzer
far-\mbox{IR} measurements, averaged along elliptical isoluminosity contours. In
Fig.~\ref{ir_vrd}b, the calculated 100\,\mbox{$\mu$m} emission distribution is
compared to the corresponding observations by \mbox{IRAS}; the low resolution of
\mbox{IRAS} prevents from seeing the wavy pattern of the spiral
arms. Figure~\ref{ir_vrd}b also shows the contributions to the
100\,\mbox{$\mu$m} flux by the warm and the cold dust. Except for the very
centre of the galaxy, contribution by the warm dust is negligible.

Using Eq.~\ref{eq:dustcoldens}, we estimated the ratio of the integral cold dust
column density to the warm dust column density to be $(7\pm1)\cdot10^3$. Since
the space density distribution of dust is proportional to the column density,
the given value is a direct estimate of the ratio of the cold dust mass to the
warm dust mass. According to \citet{gordon06}, the mass of the cold dust
component exceeds the mass of the warm component by about $10^4$ times, close to
our estimate. The warm dust gives the highest contribution in the inner bulge
region and in the spiral arms, where up to 1\,\% and up to 0.1\,\% of the total
dust mass can be ascribed to the warm dust, respectively.  This demonstrates the
insignificance of the warm dust as an absorber of starlight. On the other hand, its
contribution has to be estimated and subtracted carefully from far-\mbox{IR} maps to
avoid an overestimation of the extinction caused by the cold dust.

The derived maps of the temperature distribution and the 100\,\mbox{$\mu$m} flux
of the cold component allowed us to calculate the total optical depth along each
line of sight (Fig.~\ref{fig:tolm3}b). Within the main body of \mbox{M\,31} the
total optical depth is $\tau_{\mathrm{max}} = 0.8$--1.2 and drops to about 0.3
at the physical limit of the currently available Spitzer far-\mbox{IR} data at
about 17\,\mbox{kpc}. Beyond that, we have assumed its smooth decrease to zero
in our model. The map of the total optical depth was used for estimating the
intrinsic extinction of \mbox{M\,31}, but it can also be used for estimating the
attenuation of background objects.

According to the model presented in Sect.~\ref{sec:dustdisc} the number density
distribution of the dust disc is represented by an axi-symmetric distribution
(\ref{eq:ntau}). In the $r$-direction, the number density distribution is
determined on the basis of the far-\mbox{IR} intensity distribution. The derived
number surface density distribution of dust (which is proportional to
$\tau_{\mathrm{max}}$) for \mbox{M\,31} is given in
Fig.~\ref{fig:tolm3}c. According to the present model, an approximation of the
dust disc with a simple exponential law using $N=1$ in Eq.~\ref{eq1} gives
$(a_0)_{\mathrm{dust}} = 17$\,\mbox{kpc} as its effective scalelength. The
scaleheight of the dust disc is set to be 1.8 times the scaleheight of the
stellar disc and is thus 0.24\,\mbox{kpc}.
\begin{figure}
  \resizebox{\hsize}{!}{\includegraphics{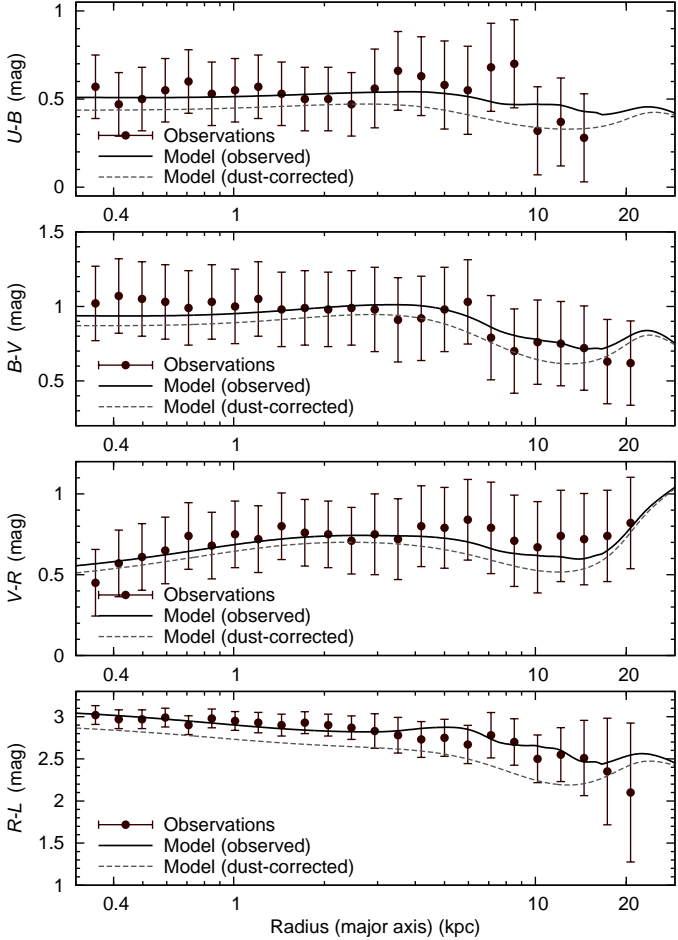}} \caption{Distribution
    of the observed (data points) and modelled (solid lines for the apparent and
    dashed lines for the dust-corrected) colour indices along the major
    semi-axis.}\label{col_major}
\end{figure}
\begin{figure}
  \resizebox{\hsize}{!}{\includegraphics{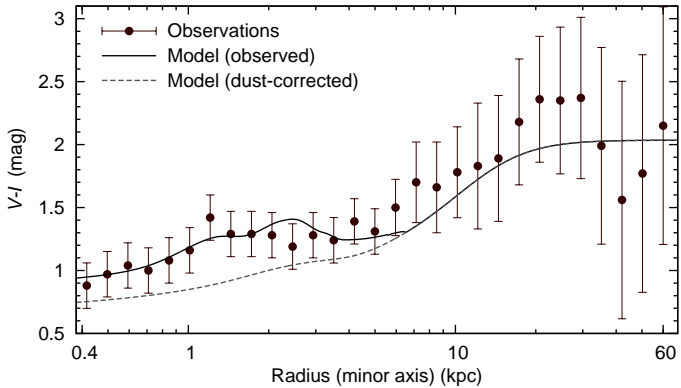}} \caption{Distribution
    of the observed (data points) and modelled (solid lines for observed and
    dashed lines for dust-corrected) $V-I$ colour index along the minor
    semi-axis.}\label{col_minor}
\end{figure}
\begin{figure}
  \resizebox{\hsize}{!}{\includegraphics{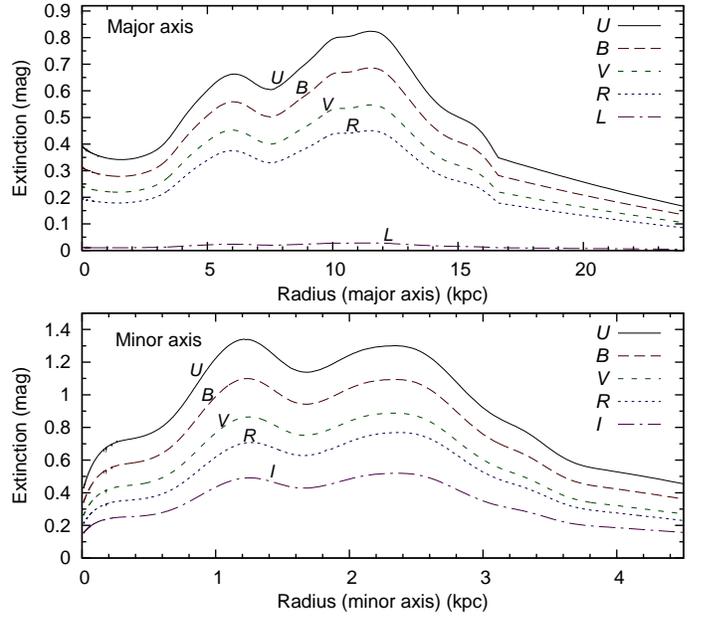}} \caption{Dust
    extinction in \mbox{M\,31} for $U$, $B$, $V$, $R$, $I$, and $L$ filters,
    calculated from the model. Upper panel: extinction along the major
    semi-axis; lower panel: extinction along the minor semi-axis.}\label{neel}
\end{figure}

The parameters of each stellar component in the best-fit model: the harmonic
mean radius $a_0$, the axial ratio $q$, the structural parameters $N$, the
dimensionless normalising constants $h$ and $k$, the relative size $\kappa$ of
the hole in the young disc component, and the absorption-corrected
$U\/B\/V\/R\/I\/L$-luminosities are given in Table~\ref{param}, together with
the uncertainty estimates. The calculated surface brightness distributions are
in good agreement with the observed surface brightness distributions.

Deviations from the observations can be further lowered by increasing the number of
components. On the other hand, due to the uncertainties of the surface
brightness data and the moderate number of available colours, the component
parameters are degenerate and the best-fitted solution is not unique: a change
of some parameters of one component can often be compensated with a change of
certain parameters of another component. Without including additional
observational data, a higher number of components would decrease the uniqueness
of the model, probably also decreasing its correspondence to the real
galaxy. The aim of the present paper was to map the effects of dust attenuation,
thus metallicity determinations in M\,31 and chemical evolution model
calculations have not been used for laying further constraints on the component
parameters. More details on the coupling of the parameters of M\,31 can be found
in \citet{tenjes94}. At present, the lack of data leaves the halo to be the most
undetermined component, especially in the $U$, $B$, $R$, and $L$
colours. Similarly, because of a slight inconsistency of the observational data
close to the centre of the galaxy, the $V$ and $R$ colours and the corresponding
colour index of the bulge cannot be determined well. The $I$ colour surface
brightness distribution is available only along the minor axis. For this reason
we cannot clearly distinguish the contributions of the disc and the young disc
to the $I$ colour and the corresponding luminosities are rather uncertain.

The axial ratio of the bulge in our best-fit model is $q=1.06\pm 0.03$,
referring to a slightly prolate shape of this component. This result seems to be
contradicting earlier estimates $q=0.6$--0.8
\citep{de-Vaucouleurs:58,Kent:89,tenjes94}. However, in those studies the
intrinsic absorption has not been considered. Extinction is stronger for a given
distance along the minor axis than for the same distance along the major axis in
the bulge region, causing a decrease of the apparent axial ratio. Without the
intrinsic absorption, our model would yield $q=0.8$ as the axial ratio of the
bulge, in agreement with previous studies.

To compare the scalelength of the dust disc with that of the stellar disc we
fitted a single exponential stellar disc to \mbox{M\,31}; the model parameters
for the bulge and halo components were kept fixes according to the
four-component approximation. The single disc has a scalelength
$(a_0)_{\mathrm{stars}} = 9.2$\,\mbox{kpc}, thus $(a_0)_{\mathrm{dust}} = 1.8~
(a_0)_{\mathrm{stars}}$, which agrees with the seven late-type galaxies studied
by \citet{Xilouris:99}.

Figure~\ref{heledus} shows the modelled surface brightness distributions along
both the major and the minor axes, together with the corresponding observational
data. The contribution of the individual components and the intrinsic,
dust-corrected surface brightness profiles are also shown. Note that while
moving from $U$ to $L$, the young disc component becomes less dominant, because
the luminosity of the star-forming ring is dominated by young blue stars. In
Figs.~\ref{col_major} and~\ref{col_minor}, the distributions of the modelled and
the observed colour indices are shown.

\begin{figure}
  \resizebox{\hsize}{!}{\includegraphics{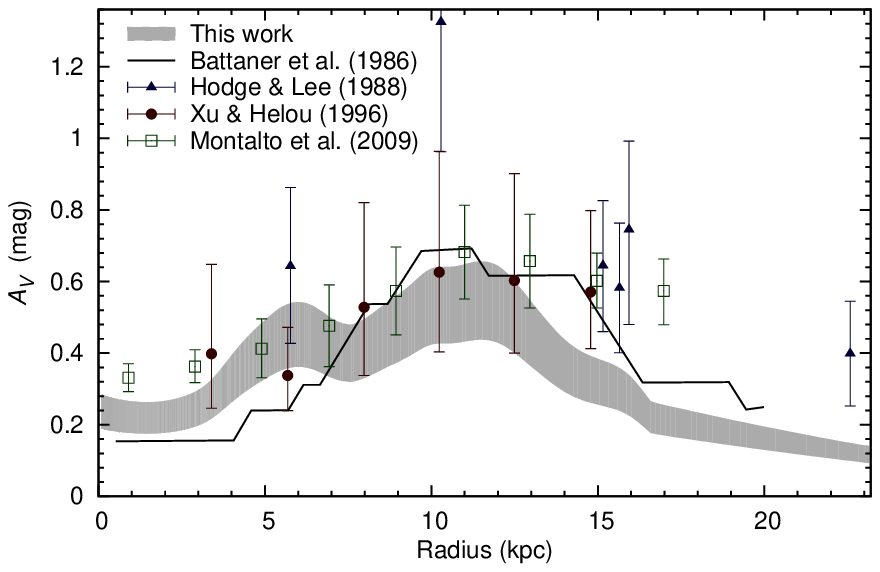}} \caption{Calculated dust
    extinction for the $V$ filter (filled area), compared to earlier estimates
    by \citet{battaner86}, \citet{Hodge:88}, \citet{xu96}, and
    \citet{Montalto:09}.}\label{neel_vrd}
\end{figure}

Figure~\ref{neel} presents the calculated dust extinction as a function of
radius for different filters. To some extent, these distributions are
model-dependent, but should be applicable with reasonable precision for
correcting any previous or coming observations of the surface brightness
distribution of \mbox{M\,31}. The largest uncertainties are related to the
derivation of the optical depth of the dust disc from the infrared
observations. The corresponding errors in Fig.~\ref{neel} would be of the order
of 20 per~cent (for the uncertainties of $A_V$ see Fig.~\ref{neel_vrd}).  Dust
extinction along the minor axis is higher than along the major axis due to
projection effects. The minor axis data represent the closer side of the galaxy,
where most of the stars lay behind the dust disc, while along the major axis,
the number of stars lying in front and behind the dust disc is equal. This
effect is also illustrated by Fig.~\ref{heledus}.

The total luminosity of \mbox{M\,31}, corrected for the intrinsic absorption is
$L_B = (3.64 \pm 0.15) \cdot 10^{10} {L_{\sun}}$, corresponding to an absolute
luminosity of $M_B=-20.89 \pm 0.04$\,\mbox{mag}. The dust disc absorbs as much
as 20\,\% (or 0.24\,\mbox{mag}) of the total $B$-flux. Roughly half of the
$B$-flux is radiated by the discs (25\,\% by the old disc and 30\,\% by the
young disc), about one quarter (22.5\,\%) by the bulge, and one quarter
(22.5\,\%) by the halo. The integrated dust-corrected intrinsic and visible
colour indices are given in Table~\ref{table:3}.
\begin{table}
  \caption{Intrinsic and visible colour indices of \mbox{M\,31}.} \label{table:3}
  \centering \begin{tabular}{lllll}
    \hline\hline
    \mbox{M\,31} & $U-B$ & $B-V$ & $V-R$ & $R-I$ \\ & (mag) & (mag)
    & (mag) & (mag) \\ \hline Intrinsic & 0.30 & 0.60 & 0.76 & 0.28\\ Visible &
    0.33 & 0.63 & 0.81 & 0.31\\
    \hline
  \end{tabular}
\end{table}

The ratio of the luminosity of the spheroidal structures (the bulge and the
halo) to the total luminosities is 0.45 in $B$ and 0.35 in $V$.

\section{Discussion} \label{sec:discussion}

\begin{figure}
  \resizebox{\hsize}{!}{\includegraphics{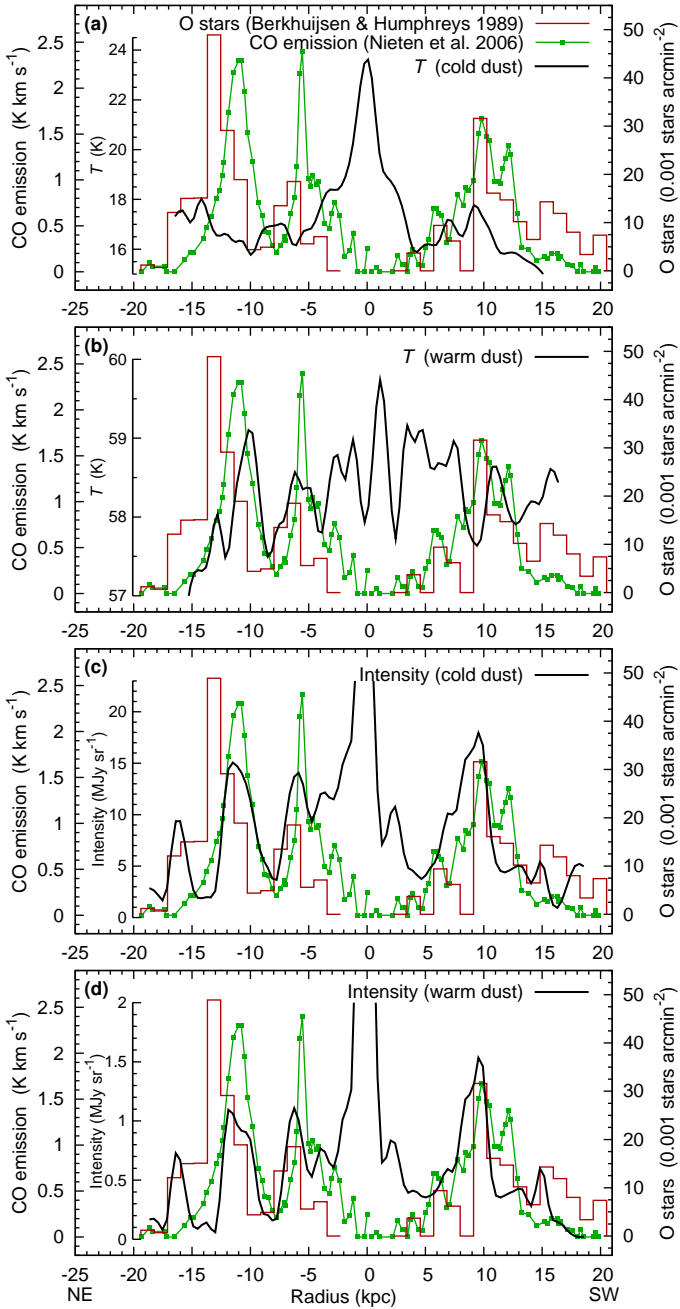}} \caption{Distribution
    of the dust properties derived from present model (black solid lines)
    together with the distributions of \mbox{CO} emission \citep[green dotted
    line,][]{Nieten:06} and young O~stars \citep[red
    histogram,][]{Berkhuijsen:89} along the major axis. (a)~-- cold dust
    temperature; (b)~-- warm dust temperature; (c)~-- cold dust emission
    intensity at 100\,\mbox{$\mu$m}; (d)~-- warm dust emission intensity at
    100\,\mbox{$\mu$m}.}\label{fig:dustline}
\end{figure}

In Fig.~\ref{neel_vrd} we display a comparison of the derived dust extinction
profile to some earlier results. Assuming a constant dust to gas ratio,
\citet{battaner86} have used the \ion{H}{i} distribution to estimate visual
extinction $A_V$ inside \mbox{M\,31}. Considering the differences in methodology
and in the availability of data, their estimate matches well with our extinction
profile. \citet{Hodge:88} have used the $U-B$ versus $B-V$ colour diagrams to
determine reddening inside the disc of the galaxy. If converted into extinction,
these values remain somewhat higher than our estimates, especially the
10\,\mbox{kpc} measurement, which is off by 0.7 magnitudes; however, also the
errorbars are large. \citet{xu96} have estimated the visual optical depth using
the \mbox{IRAS} satellite 60 and 100\,\mbox{$\mu$m} maps, an ultra-violet map,
\ion{H}{i} map, and a model of heating and cooling of dust grains. Converting
their optical depth estimates to extinction yields a slightly higher level of
extinction than calculated from our model, but \citet{xu96} have also shown that
the calculated optical depth has a large scatter across the galaxy. The most
recent analysis of dust inside \mbox{M\,31} was done by \citet{Montalto:09}. In
their study, the Spitzer data have been processed in a way quite similar to
ours. In both studies the main contributor to the far-\mbox{IR} emission is the
diffuse dust. Also, the location of the peak attenuation is the same,
10--12\,\mbox{kpc} from the centre. According to our calculations, the
attenuation decreases towards the outer regions faster than towards the inner
regions.

\citet{barmby00} and \citet{fan08} have measured the mean reddening of globular
clusters of \mbox{M\,31} caused by the dust of the galaxy to be $E(B-V) = 0.16$
and $E(B-V) = 0.28$, respectively. The former value (corresponding to $A_V=0.5$)
agrees with our estimates, whereas \citet{fan08} have derived a somewhat higher
extinction ($A_V=0.87$) than suggested by our model, but once again, the scatter
of these reddening estimates is large.

In Fig.~\ref{fig:dustline} we compare our results for the dust temperature and
intensity distributions with distributions of molecular gas and young stars. The
molecular gas distribution has been measured as the intensity of the \mbox{CO}
emission by \citet{Nieten:06}; the surface density distribution of a sample of
stars with $U-V \le -0.9$ from \citet{Berkhuijsen:89} is used as the
distribution of young stars. To enable a straightforward comparison, the dust
parameters have been measured within a narrow stripe along the major axis,
similarly to the measurements of the distributions of gas and stars. No
significant correlation is seen between the cold dust temperature and the
molecular gas or the young stars (Fig.~\ref{fig:dustline}a). The temperature of
the cold dust seems to be determined by the general radiation field of the old
stars. However, a substantial amount of the emission by the cold diffuse dust
can also be powered by non-ionising ultra violet photons coming from young
stars, as shown by \citet{Popescu:00} for the case of \mbox{NGC\,891}.  On the
other hand, a correlation exists between the amount (intensity) of the cold dust
and the distribution of the molecular gas (Fig.~\ref{fig:dustline}c). A
difference is only seen at the centre of \mbox{M\,31} where the intensity of the
cold dust emission is high while almost no gas is detected. A peak at the centre
is also seen in the intensity distribution of the warm dust
(Fig.~\ref{fig:dustline}d). Correlation between the intensity and temperature
distributions of warm dust and young O~stars is moderate only
(Figs.~\ref{fig:dustline}b and~\ref{fig:dustline}d).  The offset can be caused
by the tendency to see \mbox{CO} emission from the trailing edge of the spiral
arm, while most of the detectable 24\,\mbox{$\mu$m} emission (radiated by the
warm dust) is coming from the leading edge.
\begin{figure}
  \resizebox{\hsize}{!}{\includegraphics{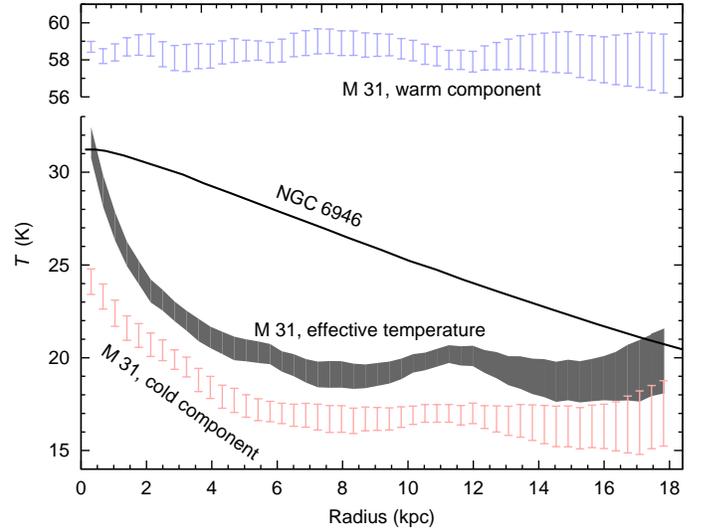}} \caption{Temperature
    distribution of \mbox{M\,31}, derived by fitting ellipses to the temperature
    maps. Dark grey area~-- effective temperature; blue and red errorbars~--
    warm and cold temperature component. The black solid line is temperature
    derived by \citet{Bianchi:00} for \mbox{NGC\,6946}. Temperature distribution
    of \mbox{NGC\,6946} is scaled to \mbox{M\,31} by taking different diameters
    of these galaxies into account.} \label{fig:dust_ellipse}
\end{figure}

We have modelled the far-\mbox{IR} spectral energy distribution as a
superposition of two modified Planck functions along each line of
sight. However, we do not actually know whether the two dust components are
physically distinctive or perhaps simply a mathematical description of a single
dust disc with higher temperature variations. To provide a description for the
case of a single dust component, we have calculated the distribution of the mean
intensity-weighted temperature of the dust. In Fig.~\ref{fig:dust_ellipse} the
mean temperature is given as a function of radius. It is seen that the
temperature decreases rather smoothly from $T \sim 32$\,K at the centre to $T
\sim 20$\,K at $R\sim 7$\,\mbox{kpc} and outwards. For illustration,
Fig.~\ref{fig:dust_ellipse} also shows the dust temperature distribution for the
Scd galaxy \mbox{NGC\,6946}, determined using a radiative transfer model by
\citet{Bianchi:00}; for this plot the radii of \mbox{NGC\,6946} have been
multiplied by a factor of 5.9 according to the ratio of the half-light radii of
these galaxies. It is seen that although the dust temperature in these galaxies
varies within the same range, it decreases more steeply in \mbox{M\,31}
indicating a more vivid bulge-disc distinction.
\begin{figure}
  \resizebox{\hsize}{!}{\includegraphics{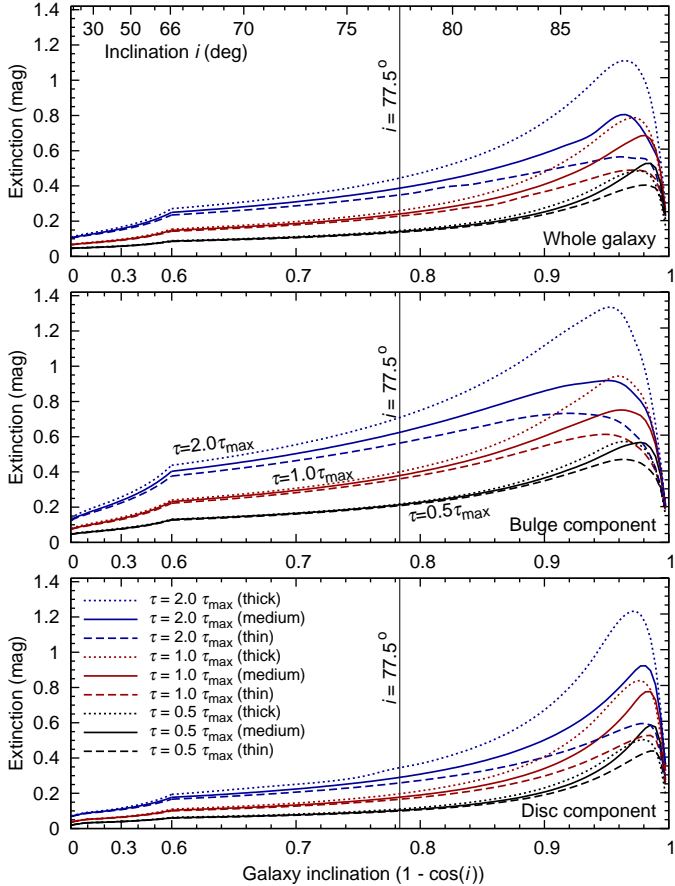}} \caption{Integrated
    dust extinction inside \mbox{M\,31} in $V$-flux as a function of the
    inclination of the galaxy for a variety of optical depths and dust disc
    thicknesses. $\tau_{\mathrm{max}}$ is the optical depth used in our
    model. Upper panel shows the extinction for the whole galaxy, middle panel
    for a bulge component, and lower panel for a disc-like component. The
    vertical line represents the inclination angle of
    \mbox{M\,31}.}\label{neel_mod}
\end{figure}

With the help of the presented model, we can also study more general
dependencies of dust extinction on the inclination angle, the total optical
depth and the dust disc thickness of a galaxy. In Fig.~\ref{neel_mod} extinction
calculations are presented for three dust disc relative thicknesses (the axial
ratios $q$): 0.005, 0.014, and 0.042, designated as `thin', `medium', and
`thick', respectively; and for three values of the optical depth $\tau$:
$0.5\tau_{\mathrm{max}}$, $1.0\tau_{\mathrm{max}}$, and
$2.0\tau_{\mathrm{max}}$, where $\tau_{\mathrm{max}}$ is the optical depth as
derived for \mbox{M\,31}. The other parameters have been set according to the
model derived for \mbox{M\,31}. The vertical line in Fig.~\ref{neel_mod} refers
to the inclination of \mbox{M\,31}. In the upper panel, the extinction is
calculated for the whole galaxy, in the middle panel for a pure bulge component,
and in the lower panel for a pure disc component.

In the case of the above-described model of \mbox{M\,31}, the total extinction
would be maximal if the inclination angle of the galaxy were approximately
88$\degr$.  At lower inclination angles (a more face-on orientation) the total
extinction decreases as a result of the decreasing line-of-sight optical
depth. For an edge-on galaxy ($1-\cos(i)=1$), the visible area of the dust disc
becomes negligible, and the total extinction is low. It is seen from
Fig.~\ref{neel_mod} that extinction is lower for the disc component than for the
bulge and increases more rapidly while moving to higher inclination values. The
extinction maximum occurs at higher inclinations for a disc than for a
bulge. Similar dependence of dust attenuation on inclination was predicted by
\citet{Tuffs:04} and are now confirmed observationally by \citet{Driver:07}
using different models and data.

Figure~\ref{neel_mod} also shows that the extinction maximum shifts slightly
towards the lower inclination angles as the optical depth increases; of course,
the total extinction increases as well. On the other hand, the total extinction
is almost insensitive to the dust disc thickness at low and intermediate
inclination angles. This is expected, since the extinction of light emitted
outside the dust disc only depends on the optical depth of the dust disc and the
dust disc geometry only affects the extinction of light emitted inside the dust
disc. In most cases, the line-of-sight thickness of the dust disc is smaller
than that of the stellar components and therefore the total extinction does not change
when changing the dust disc geometry.

In cases of very high inclination angles, the dust disc thickness becomes much
more important, especially if the optical depth is higher as well. However,
already for \mbox{M\,31} and galaxies at lower inclination angles, the
line-of-sight optical depth is the dominant factor.

The dust model used in the present paper involves certain
simplifications. Firstly, dust temperatures have not been determined on the
basis of a complete radiative transfer model. We have assumed that the
distribution of the dust temperature along each line of sight consist of a cold
component and of a warm component. The temperature maps have been derived by
determining the corresponding four parameters (the temperatures and the
intensities of the two components) according to four observational data points
for each line of sight. Although mathematically correct, the derived
temperatures may be uncertain since the chi-squared approximation is not
overdetermined. Secondly, because we do not use a radiative transfer model, the
radiation scattering is also ignored. Thirdly, our model does not consider
possible small-scale clumpiness of the dust. Although the density behaviour of
the dust disc does not follow a simple smooth exponential decrease, having a
more complicated structure thanks to the empirically determined correction
function $f_{\mathrm{dust}}(r)$ (see Eq.~\ref{eq:ntau}), possible clumps of dust
smaller than the resolution of the Spitzer far-\mbox{IR} imaging may be
present. The presence of small dust clumps would lead to a lower level of
extinction than estimated by our model, since the number of unaffected stars
becomes larger.

The vertical thickness of a realistic disc with a spiral pattern should be
varying, but the thickness of the old and the young stellar disc is assumed to
decrease monotonically in the present model.  However, the density distribution
of the young stellar disc has been modelled as a ring-like structure with a
central density minimum. Since the young disc is thinner than the old disc,
their interplay gives a slightly varying total effective thickness for the
composite stellar disc. As a result, the ratio of the stellar disc thickness to
the dust disc thickness is also variable, being larger in the regions dominated
by the old disc and smaller in the other regions. A more detailed study of the
structure of the spiral arms remains beyond the scope of the present paper.

The intrinsic luminosity and colour distribution can also be used to derive the
visible mass distribution of galaxies. Supplemented with a dynamical model
\citep[e.g.][]{Tempel:06} and kinematical observations, the visible mass
distribution can be used for deriving the distribution of dark matter in
galaxies. We will take a closer look at the distribution of visible and dark
matter in \mbox{M\,31} in a forthcoming paper.

\begin{acknowledgements}
  We would like to thank the anonymous referee for many constructive suggestions
  and comments that have helped to improve the manuscript. We thank Dr.~J.~Pelt
  for useful hints for improving the fitting process in our photometrical model
  and for estimating the errors for the parameters of the model. We are indebted
  to Dr.~Karl~Gordon, the PI of the Spitzer far-\mbox{IR} observations of
  \mbox{M\,31}. We have also benefited from communications with
  Dr.~Pauline~Barmby and Dr.~Martin~Haas.  The Spitzer Science Center and the
  \mbox{NASA/IPAC} Infrared Science Archive have been of great help in
  retrieving and processing the infrared data and we are grateful to the team
  members of these institutions. We acknowledge the financial support from the
  Estonian Science Foundation {grants 7115, 7146, 7765} and the project
  \mbox{SF0060067s08}. All the figures have been made with the gnuplot plotting
  utility. Many of the time-consuming computations for this work have been
  performed at the computer cluster of the University of Tartu.
\end{acknowledgements}

\end{document}